\documentclass[12pt,preprint]{aastex} 

\def \chandra {{\it Chandra} \/} 
\def\gta{ \lower .75ex \hbox{$\sim$} \llap{\raise .27ex \hbox{$>$}} } 
\def\lta{ \lower .75ex\hbox{$\sim$} \llap{\raise .27ex \hbox{$<$}} }

\begin{document} 
 
\title{Jets from Sub-Parsec to Kiloparsec Scales: A Physical Connection} 
 
\author{F. Tavecchio\footnote{Present address: Osservatorio Astronomico 
di Brera-Merate, via Bianchi 46, 23807 Merate, Italy}  \& L. Maraschi} 
\affil{Osservatorio Astronomico di Brera, Via Brera 28, 20121, Milano, Italy.}

\author{R. M. Sambruna} \affil{George Mason University, Fairfax, VA 22030}

\author{C. M. Urry} \affil{Yale University, Dept. of Astronomy, New
Haven, CT 06520}

\author{C. C. Cheung} \affil{Brandeis University, Department of
Physics, Waltham, MA 02454}

\author{J. K. Gambill} \affil{George Mason University, School of
Computational Sciences, 4400 University Drive, Fairfax, VA 22030}

\author{R. Scarpa} \affil{European Southern Observatory, Avenida Alonso
de Cordova 3107, Vitacura, Casilla 19001, Santiago 19, Chile}

\begin{abstract}
The \chandra discovery of bright X-ray emission from kpc-scale jets
allows insight into the physical parameters of the jet flow at
large scale. 
At the opposite extreme, extensive studies of the inner relativistic
jets in Blazars with multiwavelength observations, yield comparable 
information on sub-parsec scales. In the framework of  simple
radiation models for the emission regions we  compare the
physical parameters of jets on these two very different scales in
the only two well studied  Blazars for which 
large-scale emission 
has been resolved by \chandra. Notably, we find that the relativistic
Doppler factors and powers derived independently at the two 
scales are consistent, suggesting
that the jet does not suffer severe deceleration or dissipation. 
Moreover the internal equipartition pressures in the inner
jet and in the external X-ray bright knots 
scale inversely with  the jet cross section as expected in the
simple picture of a freely expanding jet in equipartition.
\end{abstract} 

\section{Introduction} 
 
The study of extragalactic jets has been renewed recently by the
\chandra discovery of numerous jets bright in X-rays on kiloparsec and
larger scales (Chartas et al. 2000; Worrall et al. 2001; Siemiginowska
et al. 2002; Sambruna et al. 2002).  In powerful sources, the X-rays
from the extended jet constitute a spectral component separate from
the radio-to-optical synchrotron emission. The latter can be
interpreted as inverse-Compton (IC) scattered cosmic microwave
background (CMB) photons, produced by the same population of
relativistic electrons which emit the radio to otical synchrotron
radiation (Tavecchio et al. 2000a, Celotti et al. 2001; but see Dermer
\& Atoyan 2002 and Stawarz \& Ostrowski 2002 for alternatives). The
model requires that these X-ray bright jets are still relativistic,
with bulk Lorentz factors $5-10$, at distances $\gtrsim 100$ kpc.
Fitting the two spectral components with this kind of model and the
additional hypothesis of equipartition constrains the main physical
quantities of the jet, including the Doppler factor, the magnetic
field, the density, energy distribution and, notably, the minimum
Lorentz factor of the emitting relativistic electrons, $\gamma _{\rm
min}$.
 
On much smaller scales, the innermost regions of jets in the brightest
blazars have been extensively studied through multifrequency
observations (e.g., Maraschi \& Tavecchio 2001).  The double-humped
radio-to-$\gamma$-ray spectral energy distributions of these sources
can be modeled as synchrotron plus inverse-Compton emission, yielding
robust estimates of the basic physical quantities of the emission
region close to the central black hole (e.g., Ghisellini et al. 1998,
Kubo et al. 1998, Sikora \& Madejski 2000, Sikora 2001).  From this
type of models the jet power close to the central black hole was
estimated (Ghisellini \& Celotti 2002, Maraschi \& Tavecchio 2003
(hereafter MT03)).

Comparing the physical state of the plasma {\it in the same jet} on
subparsec and kiloparsec scales can offer an important new window on
the propagation of jets, as they expand through the broad-line region,
the host galaxy, and into the intergalactic medium (e.g., Begelman,
Blandford \& Rees 1984, Bicknell 1994).  As a first step in this
direction, we discuss here such a comparison for the only two blazars,
1510-089 ($z=0.361$) and 1641+399 ($z=0.591$), for which data are
available for both the inner and extended regions of the jet: the
blazar cores are well studied observationally and were modelled by
Tavecchio et al. (2000b), (2002); X-ray and optical emission from the
large scale jets of both blazars was measured in the recent survey
with \chandra and HST of bright radio jets (Sambruna et al. 2002,
Sambruna et al. 2004 (S04)). The two sources were
included in the survey because they satisfied the selection criteria
of having radio jets of appropriate brightness and size, irrespective
of their previously known blazar properties. The plan of the paper is
as follows: in Section 2 we summarise the modeling on both scales, in
Section 3 we compare the physical parameters derived independently on
the two scales; results are discussed in Section 4. Throughout the
paper we assume $H_0=70$ km s$^{-1}$ Mpc$^{-1}$ and $q_0=0.5$.
 
\section{Modeling Jets on Small and Large Scales}

\subsection{The Inner Jet Model} 
 
The spectral energy distributions (SEDs) of the unresolved blazar cores
were modelled as synchrotron plus inverse-Compton emission, allowing for
both synchrotron and external photons as seeds for the inverse-Compton
process. The energy spectrum of the relativistic electrons is assumed to
be a broken power law with indices $n_1<3$ and $n_2>3$.  A complete
discussion of this model is given by Maraschi \& Tavecchio (2003,
hereafter MT03) and references therein.  In the homogeneous case the
model uniquely determines the main physical quantities (magnetic field,
electron density and energy, size, Doppler factor) if the spectral shapes
around the peaks of the two spectral components are observationally
determined and an upper limit to the size of the emitting region is
derived from time variability. 
The absence of a spectral break between soft and hard X-rays indicates
$\gamma _{\rm min} \sim 1$ (e.g., Tavecchio et al. 2000b). The model
 parameters derived in MT03 for the two sources are reported in Table 1.
Although not assumed in the modelling, we find a posteriori
that the emitting regions are close to equipartition.

While $\delta $ is a direct outcome of the radiative model,
the bulk Lorentz factor of the flow, $\Gamma $, depends on the
viewing angle, $\theta $. For the most
probable viewing angle, $\theta = 1/\delta $, which is the maximum
 angle allowed by the given Doppler factor, $\Gamma = \delta$.
In addition we report in Table 1 the value of $\Gamma $ corresponding
to a smaller (less probable) viewing angle $\theta = 1/2\delta $. 

\subsection{The Outer Jet Model} 

The large scale X-ray/optical radio data for the two sources are presented
and discussed in S04.   
Both jets exhibit a knotty X-ray morphology. For our analysis we used
radio, optical, and X-ray fluxes for the first knot well separated 
from the nucleus in
the \chandra image: knot B and A for PKS 1510-089 and 1641+399
respectively, at projected distances of 2.9 and 2.7 arcsec (11.8
and 13.5~kpc) from the cores.
The X-ray knots are unresolved by \chandra: their angular radii were
fixed at 1$^{\prime\prime}$, which represents an upper limit to the
actual dimension.  Radio and optical fluxes were extracted over the same
area (see S04).  The associated radio knots however appear
to be resolved in high resolution maps at the 0.5 $^{\prime\prime}$ level
(Cheung et al., in prep) thus the adopted radius should not be far from
the real one. The present analysis refers to a homogeneous
approximation.  The possible existence of strong inhomogeneities,
advanced in Tavecchio, Ghisellini \& Celotti (2003), but questioned in
Stawarz et al. (2004), would affect the results.

The SEDs of the knots are shown in Fig.~1. The data, even though sparse
in wavelength coverage, clearly indicate the presence of two emission
components, as in the prototypical case of IC/CMB jets, PKS 0637-052
(Tavecchio et al. 2000a). As in that case we modeled each SED with a
synchrotron plus IC/CMB model, assuming for the emitting electrons a
single power-law energy distribution $N(\gamma)= K\gamma ^{-n}$ between
$\gamma _{\rm min}$ and $\gamma _{\rm max}$.
The latter assumption differs from that adopted for the blazar cores
(broken power law). Due to the very limited spectral coverage the 
higher energy part of a broken power law would be 
underconstrained here. The comparison of the large scale and small
scale jet is still meaningful since the single power law adopted here 
closely corresponds to the lower energy
branch of the broken power law.

The slope of the radio spectrum of the knots is $\alpha _{r}=0.7-0.8$,
which implies $n=2.4-2.6$. In both cases the X-ray slope (0.81$\pm$
0.62, 0.66$\pm$0.86) is consistent with the radio slope within the
large errors (S04).  If, additionally, equipartition between the
magnetic and electron energy densities is assumed, the observed fluxes
provide a unique value for the physical parameters $K$, $B$, and for
the Doppler factor, $\delta$, for a fixed size of the emitting region
(Tavecchio et al. 2000a).

In the absence of information on the spectral shape in the optical
band the optical flux (in the case of PKS 1510-089 only an upper limit
is available) can be attributed either to the synchrotron or to the IC
component. Here we have chosen the first alternative for 1641+399 and
the second for PKS 1510-089 respectively.  In any case the weakness of
the optical flux constrains the minimum Lorentz factor of the emitting
electrons.  Given the steepness of the electron energy distribution
the latter quantity determines the total electron energy density.
Generally not measurable with classical radio observations (because of
self-absorption), $\gamma_{\rm min}$ is important for the derivation
of the kinetic power of the jet.  In the present two cases,
$\gamma_{\rm min}$ must be less than $\sim 10$ in order to reproduce
the observed X-ray flux and slope but larger than a few in order not
to overpredict the optical flux.  This is similar to the cases of the
four jets reported in Sambruna et al. (2002) and other jets in the
survey of S04, for which $\gamma_{\rm min}$ falls in the range 5-10.

The spectral models for the multifrequency emission from the two knots
computed with the equipartition assumption are shown in Fig.~1. The
corresponding model parameters are reported in Table 1.


\section{The Connection Between Small and Large Scales}

In the following we discuss the inner and outer jet "connection"
with regard to the bulk velocity, the transported (kinetic) power
and the internal energy density/pressure. 

\subsection { \it Bulk motion}

A comparison of the parameters independently derived for the inner and
outer jet (see Table 1) shows that the values of the beaming factors
($\delta$) for the two widely separated regions are similar. As
mentioned above a determination of the bulk Lorentz factor $\Gamma$
requires an assumption about the viewing angle.  Starting from the
largest possible angle for a given value of $\delta$, $\theta _{\rm
max}\sim 1/\delta$ for which $\Gamma = \delta$ the minimum value of
$\Gamma$ is $\delta /2$ for $\theta=0$. Since $\theta=0$ is unlikely
and unpractical, for instance for deprojecting, we report in Table 1
the range in $\Gamma$ corresponding to viewing angles between $\theta
_{\rm max}$ and $\theta _{\rm max}/2$. Note that the value of $\Gamma$
for $\theta _{\rm max}/2$ is very close to $\delta /2$, corresponding 
to $\theta=0$. We discard the possibility of
{\it much larger} values of $\Gamma$ which are also in principle
allowed for angles smaller than $\theta _{\rm max}$ but appear
unreasonable in view of the higher implied kinetic powers (MT03).

That powerful jets remain relativistic at large scales was anticipated
long ago on the basis of theoretical expectations (Blandford \& Rees
1974) and observational evidence of jet one-sidedness and
depolarization asymmetry (Laing 1988; Garrington \& Conway 1991). From
our analysis (summarized in Table 1) we derive the values of the
Doppler beaming factors at subpc and 100 kpc scales finding no
significant difference. Although some deceleration cannot be excluded,
the results do not suggest or require it.  Recent numerical
simulations for highly relativistic jets are in fact consistent with
these conclusions (Scheck et al. 2002).

\subsection { \it  Kinetic power}

An important global quantity that can connect the jet at different scales
is the transported power. Assuming that the central engine is stationary
when averaged over an appropriately long time scale ($10^5-10^6$$y$), 
the transported power should remain constant or decrease along the jet. 
To fix ideas one could refer to the "internal shock" model applicable
both to GRBs and to relativistic jets in radiosources (Spada et al. 2001). The
"instantaneous" power emitted by the central engine fluctuates but along
the propagation path the fluctuations merge and are progressively smoothed
into an almost continuous flow.
  
For both the
inner, unresolved region and the (still relativistic) resolved jet, the
transported power can be computed using the expression
\begin{equation}
P_{\rm j} = \pi R^2 \Gamma ^2 ( U^{\prime 
    }_B+U^{\prime }_e+U^{\prime }_p)c 
\end{equation}
\noindent
(Celotti \& Fabian 1993), where $R$ is the radius of a cross section
of the jet, $U^{\prime }_e$, $U^{\prime }_p$, and $U^{\prime }_B$ are
the rest frame energy densities of relativistic electrons, protons,
and magnetic field, respectively. $U^{\prime }_e$ can be expressed as
\footnote{This is the approximated version (valid for $\gamma _{\rm
min} >>1$) of the general $U^{\prime }_e=m_ec^2\int _{\gamma _{\rm
min}}^{\gamma _{\rm max}} N(\gamma) (\gamma -1) d\gamma$.} $U^{\prime
}_e=n_e<\gamma >m_ec^2$, where $n_e$ is the total electron density and
$<\gamma > $ is the average Lorentz factor.  

In general there is little direct information about $U^{\prime }_p$,
the ``matter content'' of jets. For powerful blazars an indirect
argument leads to assume a significant proton content.
In fact the ``intrinsic''
radiative luminosity from the core exceeds the power computed assuming an
e$^{\pm}$ plasma, which would lead to substantial deceleration of the jet
close to the nucleus (Ghisellini \& Celotti 2002, Maraschi \& Tavecchio
2003). If instead the electrons are neutralized by protons exerting a
negligible pressure, the computed power (dominated by the cold proton
component), yields for the inner jet a radiative efficiency of $\sim
10\%$. Thus in the following we adopt the latter assumption.

The derived powers are reported in Tab 1.  The range associated with
the uncertainty on the observing angle is of about a factor 4.  We
also tried to assess the reliability of our estimates by checking the
sensitivity of the derived parameters on the assumed radius. We
verified that a radius smaller by a factor 2 than assumed affects
the derived power by a factor of 2 (An approximate analytic treatment
gives $P=f^{-2\alpha/(3+\alpha)}$ where f ($>1$) is the reduction factor
of the size).

The results above are valid for equipartition models.  In order to
assess the effect of relaxing the equipartition assumption for the
outer jet we computed emission model parameters {\it without assuming
equipartition}.  Since we have to substitute the equipartition
hypothesis with some other condition we choose to fix $\Gamma $ and
compute models for various values of $\Gamma $ around the
equipartition values.  $\delta $ was derived from $\Gamma $ with a
fixed viewing angle (assumed to be equal to the value derived for the
blazar region) within the range reported in Table 1. This was done
using analytic approximate formulae for the fluxes rather than
performing direct spectral fits to the SEDs.
 
In Fig.2. we compare the powers resulting from different assumptions.
The continuous line represents the power computed applying the the
IC/CMB model for different values of $\Gamma $, without assuming
equipartition (each point along this curve implies a different model which
reproduces the observed fluxes).
The power increases rapidly at low $\Gamma $ because of the larger 
rest-frame energy densities required by the weaker beaming. 
For both sources it has a broad minimum in the range of  $\Gamma $s
allowed in Table 1. 

The equipartition condition (determined only by the synchrotron radio
flux) is shown by the dashed line (calculated assuming a fixed value
$\gamma_{\rm min}=10$). The intersection of the continuous and dashed
line marks the ``equipartition'' solution which corresponds
(approximately) to the parameters of Table 1 and to the spectral models
shown in Fig 1.

For comparison the dotted line shows the power
required if X-rays were emitted via the synchrotron self-Compton
process (for which the only seed photons are the synchrotron ones).
For that model, as noted by Schwartz et al. (2000) for PKS 0637-052, the
required power would be extremely large and equipartition far from
satisfied.

It is noteworthy that the kinetic power of the outer jet, for an
IC/CMB model near equipartition, is  close to the power estimated
{\it independently} for the inner jet from the blazar SED (shown
Fig.~2 by the thick horizontal dashed line; see also Table 1). 

\subsection {\it Energy densities and pressure}

The internal jet pressure (given by $p=U^{\prime }_B/3+U^{\prime }_e/3$)
in the inner, unresolved jet region can be compared with that estimated
in the outer knots. Remarkably the ratio of the inner and outer
pressures ($20 - 6 \times 10^{10}$) scales approximately with an inverse
square law with respect to the (rest-frame) size of the emission region
$R $ ($ 5 - 4 \times 10^5$).

Assuming the inner region lies at $r\sim 0.1 \, {\rm pc}\sim 3\times
10^{17}$~cm (Ghisellini \& Madau 1996), and computing the distance of
the external knots from the angular distance, deprojected with the two
values of the viewing angle, we find that the scale factors for $R $
and $r$ are similar within 1 order of magnitude.  Thus we can make the
hypothesis that the jet is almost conical over 6 orders of magnitude
in scale.

Moreover the inner and outer models independently indicate near
equipartition. Thus our results are consistent with the very simple
picture of a free jet in equipartition as described by Blandford \&
K\"{o}nigl (1979) where the magnetic field should decay with the
cross-sectional area of the jet, $A$, as $B\propto A^{-1/2}$, and the
pressure as $p\propto A^{-1}$.


The fact that the jet is free is consistent with (and supports) the
results on the conservation of jet power. The interaction with the
external medium should be weak so that only a small fraction of the
power can be dissipated through shocks. As a result the jet does not
decelerate substantially (at most a factor 2 in $\Gamma$).  An
interesting point about the jet pressure at large scale is that its
value is of the same order as the pressure of the gas inferred in the
hot haloes at comparable distances in FRI and FRII host galaxies for
which profiles has been measured (e.g., Hardcastle \& Worrall 2000,
Worrall \& Birkinshaw 2000) and in the cluster gas around some
intermediate-redshift radio-loud quasars (e.g., Hardcastle \& Worrall
1999, Crawford \& Fabian 2003). This condition could be associated
with the end of the phase of free expansion.
$p_{\rm ext}\sim 10^{-11}-10^{-12}$ erg/cm$^3$.

\section{Conclusions}

We have presented a case study of two blazars for which high-quality
radio, optical, and X-ray observations of both the small-scale and the
large-scale jet are available.  The physical parameters in the blazar
core region and in the outer knots are reasonably well --- and
independently --- determined.

A comparison between the physical quantities at the two scales 
indicates that:
 
\noindent 
1) The jet appears to maintain an almost constant relativistic
 velocity from subparsec scales to distances of hundreds of
 kiloparsecs. A deceleration of a factor 2 is allowed but not
 indicated by the derived parameters.

 \noindent
2) The transported power inferred for the outer jet is remarkably
similar to that estimated close to the nucleus.

\noindent
3) The pressure at the two scales is consistent with a simple scaling
relation with the jet cross-section, $p\propto A^{-1}$ and the jet
geometry between the two scales is approximately conical.
 

Admittedly the results are model dependent and the quantitative
validity of the derived parameters holds within factors of a few,
 except for the beaming factor $\delta$ which is better determined
due to the strong dependence of all quantities on it.
Nevertheless it is noteworthy that all the results are consistent with
the simple scenario of a freely expanding jet in equipartition.
 
While other possibilities, such as magnetic confinement
(e.g. Begelman, Blandford \& Rees 1984), can not be excluded.  the
conditions for a free expansion are certainly satisfied if we compare
the derived pressures in the jet with that of an external medium. The
bright external knots could ultimately derive from processes
associated with the end of the validity of the conditions of free
expansion.

The analysis presented here offers new elements relevant for the
understanding of the global behaviour of jets. Clearly more
observations are needed. While angular resolution, important to
investigate the issue of sizes and locations of knots in jets, is not
likely to improve significantly in the near future, deeper
observations and the study of a larger number of sources will
certainly provide a wider context for investigating the issues
addressed here.

\acknowledgements

We thank the referee for a careful and critical reading of the
paper. FT and LM acknowledge support from grant COFIN-2001028773-007
and ASI-I/R/047/02. RMS gratefully acknowledges support from an NSF
CAREER award and from the Clare Boothe Luce Program of the Henry Luce
Foundation.

\clearpage 
 
\vskip 1.5 truecm 
 
\centerline{ \bf Figure Captions} 
 
\vskip 1 truecm 
 
\figcaption[sed]{Spectral Energy Distributions for the knots of the
blazars 1510-089 and 1641+399 analyzed in this work.  For 1510-089,
only an optical upper limit could be obtained. The solid line is the
synchrotron-IC/CMB model used to reproduce the data (see text for
details).}
 
\figcaption[power]{Jet power inferred on the kiloparsec scale and in
the core for 1510-089 and 1641+399. The intersection of the
large-scale power estimated using equipartition (short-dashed line)
and that estimated with the IC/CMB model (solid line) is very close to
the core power (long dashed horizontal line).  The power required by a
synchrotron self-Compton model (dotted) is much larger, well
above the usual power estimates in these sources.}
 
\newpage 

\begin{table}
\begin{center}
\begin{tabular}{ccccccc|ccc}
\multicolumn{7}{c}{Parameters of the radiative model}&\multicolumn{3}{c}{Inferred quantities} \\
\hline
&&&&&&&&&\\
& $\gamma_{\rm min}$ & $n$ & $U^{\prime }_B$& $U^{\prime }_e$& $R$&$\delta$&$\Gamma$ & $P_{\rm j}$& $r$ \\
& & &erg/cm$^3$ &erg/cm$^3$ & cm & & & $10^{47}$erg/s& cm \\
\hline
1510-089&&&&&&&&&\\
inner & 1& 1.7& 8.9$\times10^{-2}$ & 9.5$\times10^{-2}$& 3$\times 10^{16}$ & 19 & 19-9.5 & 5-1.25&$3\times 10^{17}$\\
outer & 10& 2.7& 4.1$\times 10^{-13}$& 4.15$\times 10^{-13}$& 1.3$\times 10^{22}$ &16 & 16-8.6 & 4.4-1.3 &0.6-1.1$\times 10^{24}$\\ 
&&&&&&&&&\\
\hline
1641+399&&&&&&&&&\\ 
inner & 1& 1.5& 0.3& 0.4& 4$\times 10^{16}$& 9.7& 9.7-5& 1.2-0.3& $3\times 10^{17}$\\
outer & 10& 2.4&  5.3$\times10^{-12}$& 6.4$\times10^{-12}$& 1.5$\times 10^{22}$& 8 & 8-4.3 & 4.2-1.1& 3.2-6.5$\times 10^{23}$\\ 
&&&&&&&&&\\
\hline
\end{tabular} 
\caption{Physical parameters estimated for the inner blazar jet (first
line) and for the knot at large distance (outer jet, in the
equipartition case, second line) for the two sources. For the core
blazar region we report only the first index of the electron
distribution ($n=n_1$). $\delta $ is the Doppler factor, $R$ indicates
the radius of the (spherical) emitting region, $r$ is the distance of
the region from the central black hole. We indicated the allowed ranges of
$\Gamma $, $P_{\rm j}$ and $r$ (for the outer jet only). See text for details.}
\end{center} 
\end{table}

\begin{figure} 
\centerline{\plotone{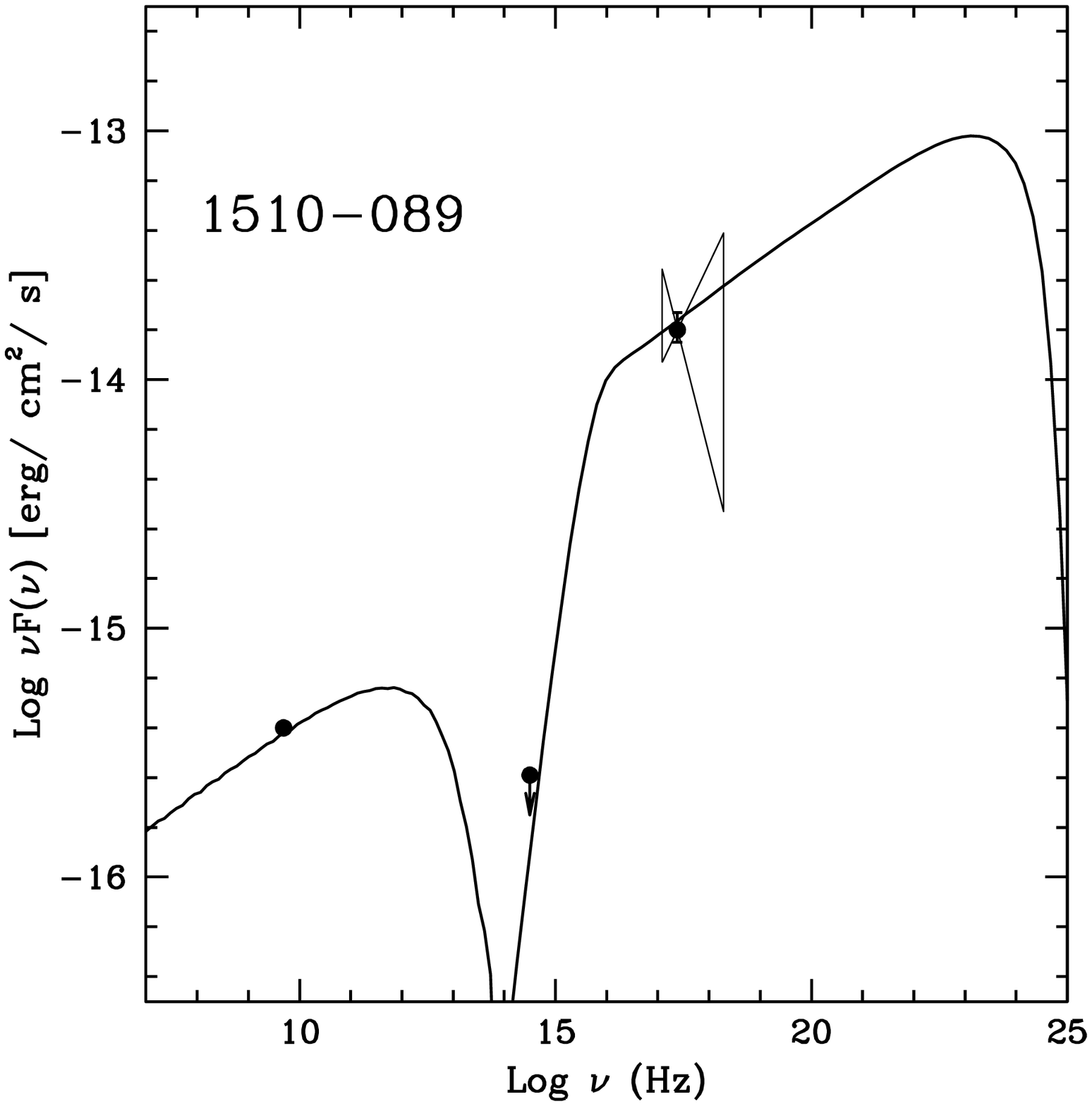}} 
\end{figure} 
 
\begin{figure} 
\centerline{\plotone{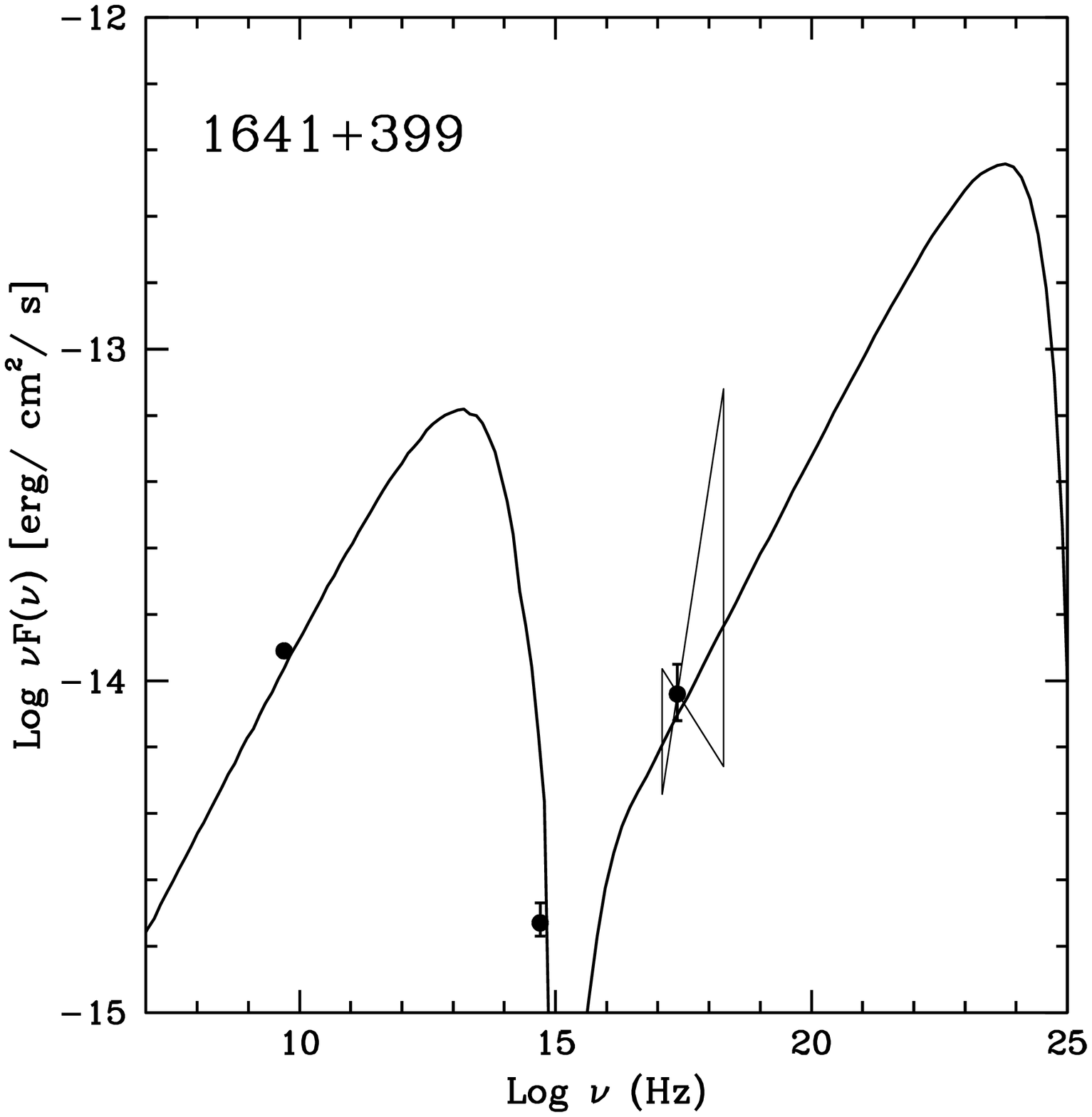}} 
\end{figure}

\begin{figure} 
\centerline{\plotone{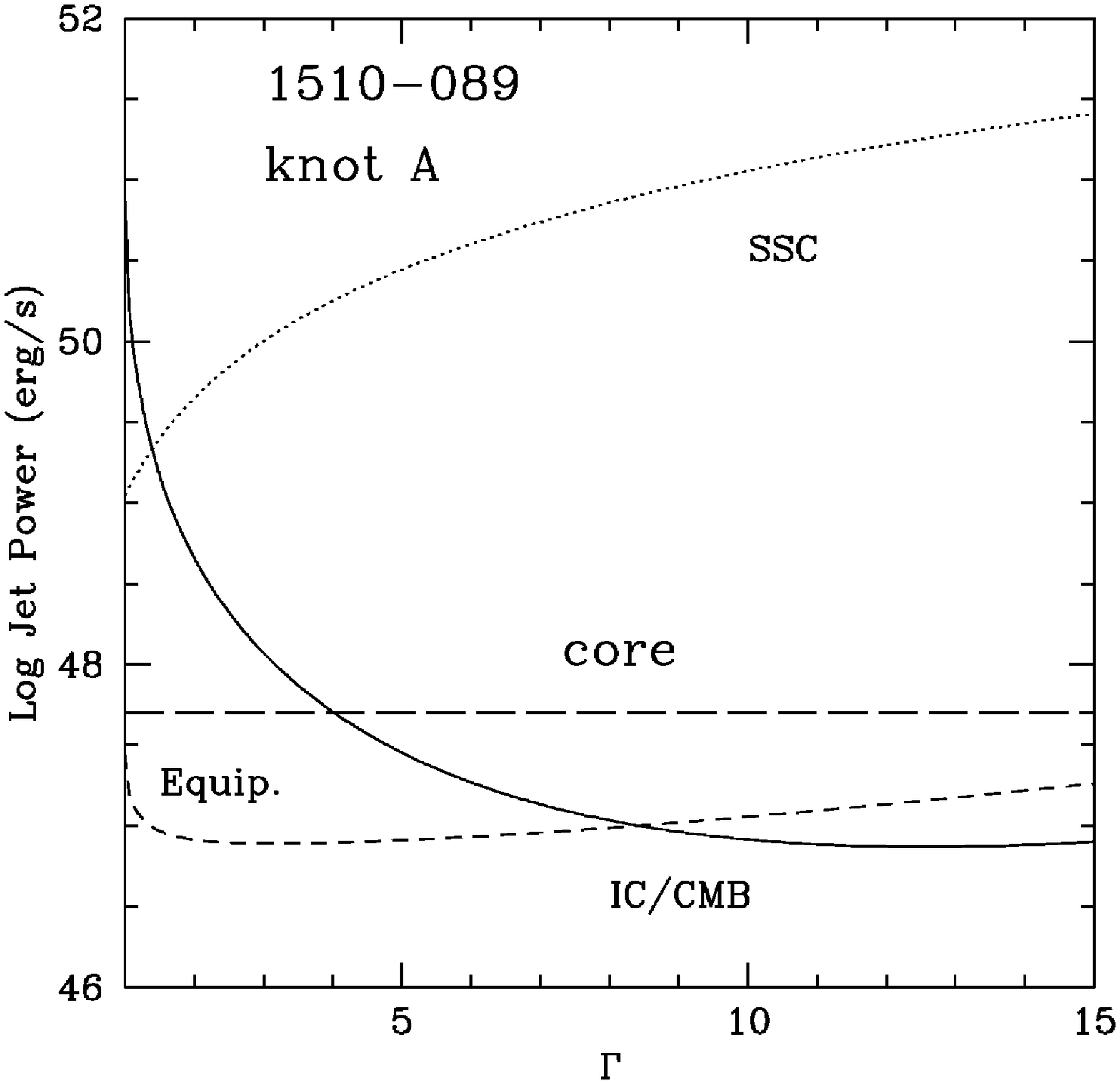}} 
\end{figure} 
 
\begin{figure} 
\centerline{\plotone{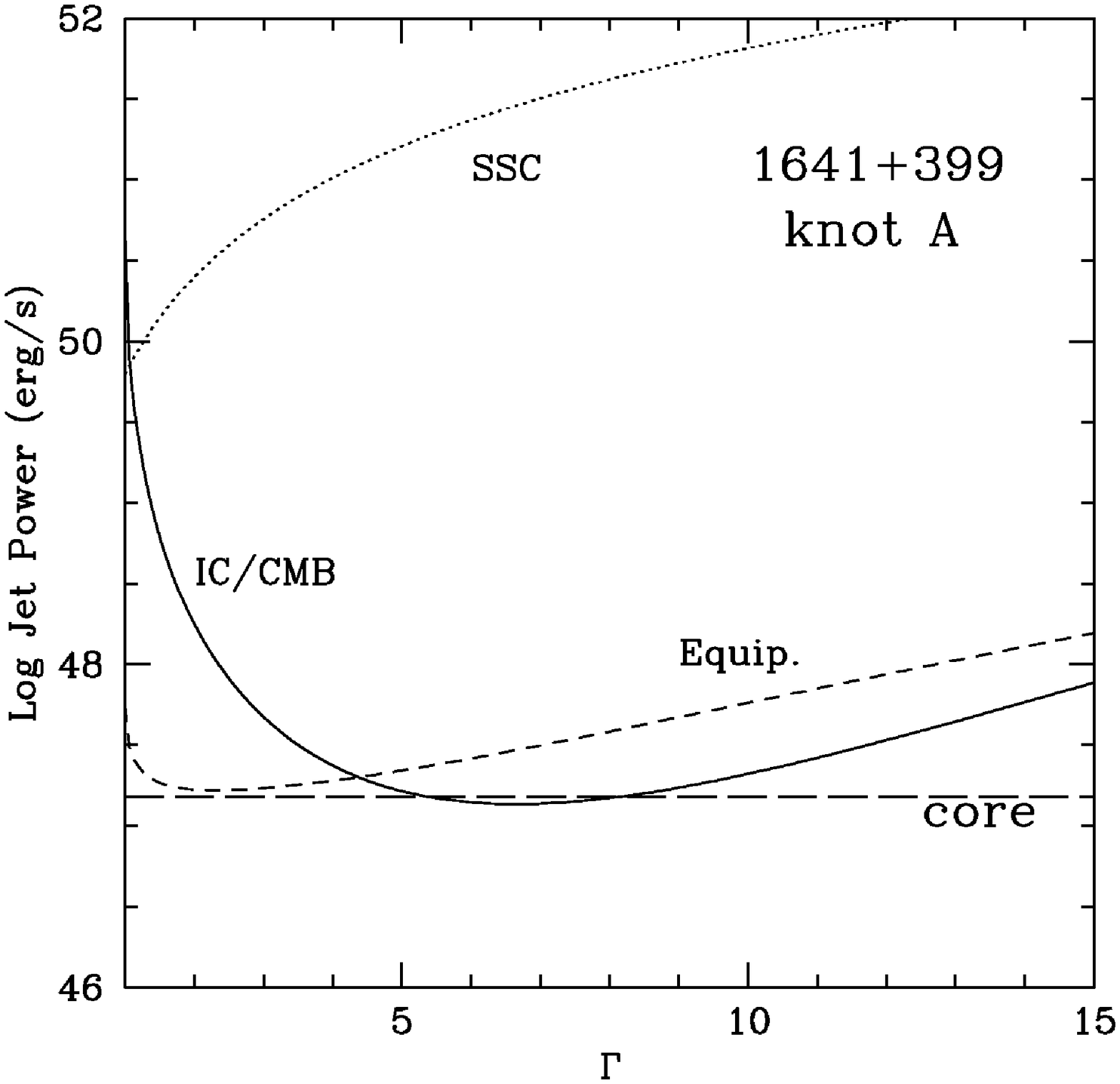}} 
\end{figure}

\end{document}